# Flexible and Scalable Data-Acquisition Using the *artdaq* Toolkit

Kurt Biery, Eric Flumerfelt, John Freeman, Wesley Ketchum, Gennadiy Lukhanin, Adam Lyon, Ron Rechenmacher, Ryan Rivera,

Lorenzo Uplegger, Margaret Votava

*Abstract*— The Real-Time Systems Engineering Department of the Scientific Computing Division at Fermilab is developing a flexible, scalable, and powerful data-acquisition (DAQ) toolkit which serves the needs of experiments from bench-top hardware tests to large high-energy physics experiments. The toolkit provides data transport and event building capabilities with the option for experimenters to inject *art* analysis code at key points in the DAQ for filtering or monitoring. The toolkit also provides configuration management, run control, and low-level hardware communication utilities. Firmware blocks for several commercial data acquisition boards are provided, allowing experimenters to approach the DAQ from a high level. A fully-functional DAQ "solution" of the toolkit is provided in *otsdaq*, sacrificing some flexibility in favor of being ready-to-use. *artdaq* is being used for several current and upcoming experiments, and will continue to be refined and expanded for use in the next generation of neutrino and muon experiments.

## I. Introduction

Data Acquisition (DAQ) for modern high-energy physics experiments must handle very high data rates and intelligently apply complex trigger algorithms in real-time. Additionally, several experiments may contribute to a single facility, and a shared data format is necessary for interoperability. The experiments are also constrained by schedule and budgetary concerns, which makes already-developed and supported DAQ solutions more desirable.

The Scientific Computing Division's Department of Real-Time Systems Engineering at Fermilab has developed the *artdaq* DAQ framework [1] to fulfill the needs of the current and upcoming neutrino and muon experiments, such as DUNE [2], SBND, ICARUS, MicroBooNE [3], and Mu2e [4]. The flexibility of the *artdaq* toolkit allows for per-experiment customizations while maintaining a set of standards that allows developers familiar with one DAQ system to quickly familiarize themselves with another *artdaq*-based DAQ. Additionally, the short-baseline neutrino program at Fermilab (SBND, ICARUS and MicroBooNE) will do cross-analyses for sterile neutrino and other new-physics searches, and the use of common tools is a valuable component for enabling such meta-analysis.

The Scientific Computing Division at Fermilab has adopted a strategy of developing common tools for use by all experiments. This allows for centralized development and easier collaboration between experiments. The *art* analysis suite [5] is an example of this as a single framework that is extremely flexible and experiment-agnostic. In both *art* and *artdaq*, the core functionality of data movement is implemented on the behalf of the user, so they can implement algorithms that simply work on data.

By incorporating *art* into the DAQ toolkit, complex trigger algorithms can be run on data in real-time, limited only by available processing power. Additionally, *art* online monitors can be attached to a running system and provided with a subset of the data for control room displays. Data are saved in *art*'s ROOT format by default, reducing the amount of preprocessing necessary for offline analysis.

*artdaq* provides considerable flexibility in the technology options that can be used for various DAQ functions, so for small experiments and users who want to get started quickly, we provide a ready-to-use instance of the toolkit called *otsdaq*. *otsdaq* makes underlying technology choices for users and allows them to get started with a small amount of configuration. The *otsdaq* instance of *artdaq* is intended to be used off-the-shelf, and as part of that we provide commonly-used firmware blocks that have been developed for a set of supported FGPA boards.

## II. ARTDAQ System Architecture

Several components make up the data path of a typical *artdaq* system (Fig. 1). A Fragment Generator is run in a BoardReader for each discrete hardware readout component. The BoardReaders send their data to a set of EventBuilders, which build events based on the sequence number of the Fragments coming from the BoardReaders. A RoutingMaster can perform load-balancing between *artdaq* processes. The DataLogger is tasked with writing data to disk, and it also sends the data stream on to a Dispatcher, which makes it available to online monitors.

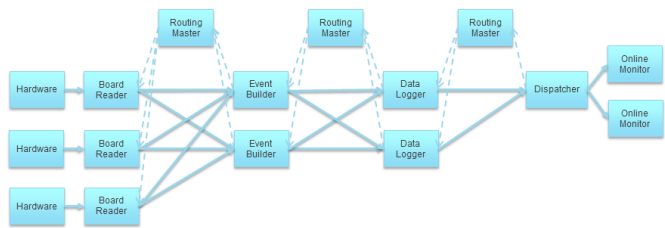

Fig. 1. *artdaq* System Layout showing data flow relationship between processes

Data transport through the *artdaq* system is accomplished using the "Fragment" class (Fig. 2). This class consists of a header containing all of the information needed by *artdaq* for routing and event building, optional metadata, and the experiment-defined payload. Data from the readout hardware is read directly into the Fragment payload and the Fragment header is written by a "Fragment Generator" class. Fragment Generators are responsible for implementing any custom readout code needed by the experiment. The BoardReader can buffer and repackage its Fragments using one of several Data Request modes, if desired. The data request modes allow a Fragment Generator to pick data from a data stream using an experiment-defined timestamp index, or integrate detector controls data that is acquired at a different rate than DAQ data.

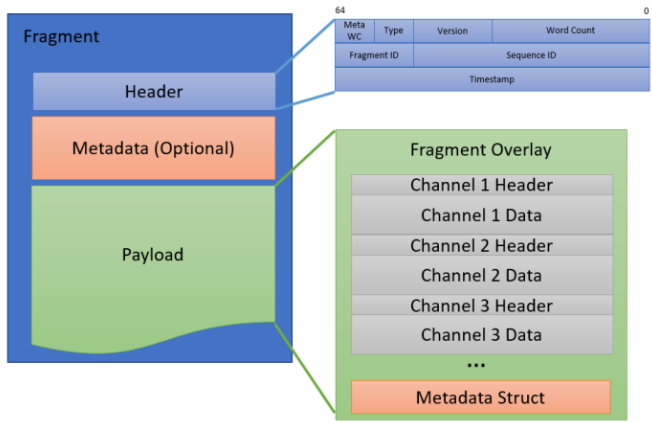

Fig. 2 Diagram of the Fragment class

Fragments are transported through the *artdaq* system using "Transfer Plugins". Several Transfer Plugins are provided with the toolkit, a Shared Memory transfer, a TCP Socket transfer, and an "Autodetect" transfer which selects Shared Memory or TCP based on whether the source and destination are on the same host (Fig. 3). Additionally, other protocols can be used by implementing the Transfer Interface. The DataSenderManager class selects the destination for each Fragment based on the sequence ID, using information from the RoutingMaster, if used. The DataReceiverManager receives data from all sources and ensures that none of the senders gets too far ahead of the others.

The data from the BoardReaders are received into a shared memory segment and built into events using the Fragment sequence ID as the key. These events are then sent to an *art* process for an optional filtering step, using experiment-defined *art* modules. The EventBuilder is also responsible for generating data request messages used for triggering readout from the BoardReaders. It does this by taking the timestamp of the first Fragment received for a given sequence ID and generating a request message which is multicast to all BoardReaders.

The RoutingMaster is responsible for creating routing tables linking sequence ID to destination using tokens sent from the Fragment receivers (EventBuilders or DataLoggers) representing available buffers. In the absence of RoutingMasters, *artdaq* uses a Round-Robin routing scheme based on the sequence ID of the Fragments. Experiments may define their own "Routing Policy" plugins which define the mapping between tokens and routing table entries. Routing tables are sent to the senders (BoardReaders or EventBuilders) via a multicast protocol that uses acknowledgements to ensure correct transmission.

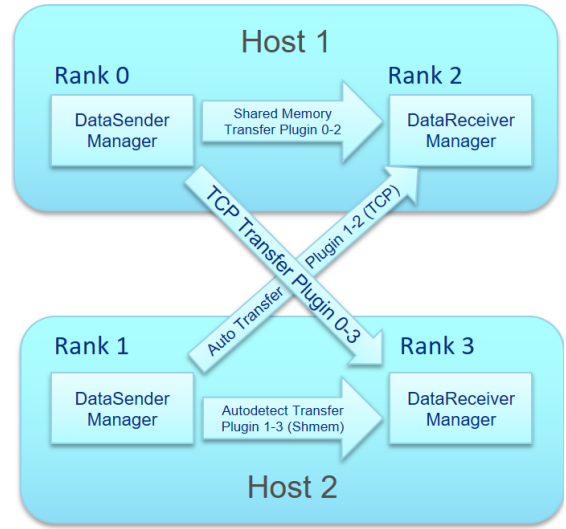

Fig. 3 Transfer Plugins in *artdaq*

Events passing the filters in the EventBuilder are sent on to the DataLogger component, which writes these data to disk. The DataLogger also sends data to the Dispatcher, however this data transfer is explicitly non-blocking and data may be dropped if the Dispatcher cannot keep up. Multiple DataLoggers may be used to distribute disk-writing load to multiple hosts. *art* workflows are supported in the DataLogger, allowing for multiple data streams or additional filtering. Additionally, the DataLogger and EventBuilder layers may be combined, with one *artdaq* process responsible for both filtering and disk-writing.

The Dispatcher forms the bridge between the DAQ system and any online monitoring the experiment wishes to perform. It supports dynamic addition and removal of online monitor processes and can run *art* filters to determine what data to send to a given monitor. This "local" filtering is intended to reduce the amount of data sent to the monitors, but can be more complex than a simple pre-scale. The online monitoring processes themselves are usually *art* instances that run reconstruction and display modules.

### III. ARTDAQ FRAMEWORK TOOLSET

In addition to the primary responsibility of data readout, the *artdaq* framework contains several tools that are necessary when building a complete DAQ system. These tools implement run control, configuration management, message logging, metric reporting and low-level hardware management. Interfaces are provided so that an individual experiment can pick and choose from the menu of available options or create their own.

*artdaq* inherits a plugin system from *art*, where it is used to dynamically load analysis modules. The plugin system handles locating and loading of shared libraries containing plugin implementations. *artdaq* provides several interfaces which allow for new functionality to be loaded at runtime. The FHiCL language [6] was developed to describe the run-time configuration of an *art* job, and *artdaq* also uses FHiCL to configure its processes. Apart from the basic readout chain, most functionality in *artdaq* is implemented using plugins.

Each experiment that uses *artdaq* creates a repository of plugin code, customized to their needs. At a minimum, this includes their custom Fragment Generator and an "Overlay", which is used to decode the binary data stored in the *artdaq* Fragment. Overlays are primarily used in filters and offline processing to interpret the payload "blob" and define the format of the optional metadata. Metadata is generated in the BoardReader and can contain conditions data, location information, or any other quantities that should be associated with the data payload.

In addition to the *art* modules used in the filtering and online monitoring, plugins can change the run control messaging protocol, the data transport mechanism between *artdaq* processes, the destinations to which messages are logged, and where metrics are reported. *artdaq* provides a library of plugins for each of these tasks, and implementations of protocols that are not experiment-specific may be promoted from experiment repositories into the *artdaq* framework itself to be made available to all users of *artdaq*.

The toolkit contains several utilities useful in a running DAQ system. A database interface to document-based backends such as MongoDB, adding a layer of change control to the set of FHiCL files used to configure *artdaq*. A web-based editor talks to the database backend and allows users to make minor changes to their configuration between runs. Each file is independently versioned, and each configuration in the database has a version number representing an immutable set of configuration files. A complete set of configuration files is extracted from the database for each run and stored in a run records directory. Additionally, the text of the configuration documents is stored in the data files.

A high-speed messaging facility, TRACE [7], logs messages to a memory buffer, and dispatches selected messages to *art*'s Message Facility, where they are routed to configurable destinations. TRACE is designed for ultra-high performance, using fixed-size memory-mapped files to store variables associated with the message to delay the expensive string formatting steps until they are needed. TRACE uses "level masks" to enable and disable selected messages within the system, allowing debugging to proceed even on "live" systems. The MessageFacility output is enabled via another mask so only selected messages are processed through this "slow" path. The artdaq MessageFacility extensions provides a destination plugin that sends messages to a viewer application, which can be used to monitor the state of the artdaq system (Fig. 4).

IV. CONCLUSION

By providing an efficient, scalable, and flexible DAQ system, we have simplified the task of supporting multiple experiments' DAQ systems and removed the burden of developing high-throughput systems from collaborations. Additionally, by supporting *art* analysis modules in the DAQ, we reduce the number of frameworks necessary for scientists to learn to make quality triggering and monitoring algorithms. The *otsdaq* product allows rapid deployment for simple DAQ systems, and it includes features such as a run control system using XDAQ [8] that allows *artdaq*-based DAQ systems to be used for development and calibration-mode runs at CMS. Our flexibility consistently allows us to support new users with novel requirements with minimal changes to the framework, and the modularity of the system allows interesting ideas to be integrated into the core framework.

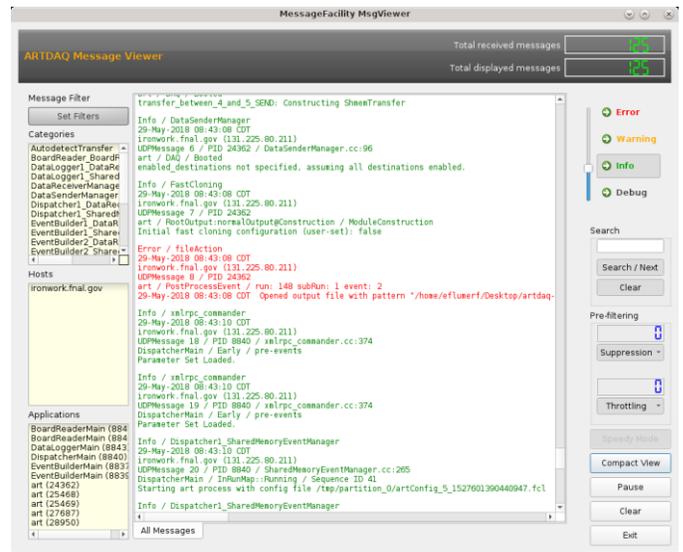

Fig. 3 The *artdaq* MessageFacility Viewer

Each experiment that uses the *artdaq* toolkit provides a new set of requirements, spurring continued refinement. For example, the software trigger of Mu2e require complex analysis to be performed in an extremely tight time budget, while protoDUNE (and eventually DUNE) will have extremely high data-volume requirements. As an integrated DAQ toolkit, the improvements made on the behalf of one experiment inform and improve all future and contemporary experiments.